# Analysis of FDML lasers with meter range coherence

Tom Pfeiffer[a], Wolfgang Draxinger[a,b], Wolfgang Wieser[b], Thomas Klein[b], Markus Petermann[b] and Robert Huber[a]

[a]Institut für Biomedizinische Optik, Universität zu Lübeck, Lübeck, Peter-Monnik-Weg 4, 23562 Lübeck, Germany; [b]Optores GmbH, Gollierstr. 70, 80339 Munich, Germany

## ABSTRACT

FDML lasers provide sweep rates in the MHz range at wide optical bandwidths, making them ideal sources for high speed OCT. Recently, at lower speed, ultralong-range swept-source OCT has been demonstrated[1, 2] using a tunable vertical cavity surface emitting laser (VCSEL) and also using a Vernier-tunable laser. These sources provide relatively high sweep rates and meter range coherence lengths. In order to achieve similar coherence, we developed an extremely well dispersion compensated Fourier Domain Mode Locked (FDML) laser, running at 3.2 MHz sweep rate and 120 nm spectral bandwidth. We demonstrate that this laser offers meter range coherence and enables volumetric long range OCT of moving objects.

## 1. INTRODUCTION

Modern FDML lasers have paved the way to high fidelity swept source OCT systems with very high A-scan rates and long coherence lengths. FDML lasers exhibit the unique feature being capable of wavelength sweep repetition rates well into the MHz range[3-8], because they do not suffer from inherent physical constraints with respect to wavelength sweep rate[9-14]. Hence, they found widespread applications from ultrafast OCT[7, 15-18], over non-destructive sensing and testing[19-23] to optical molecular[24, 25] and functional imaging[26, 27] and even short laser pulse generation[28]. The extension of the accessible wavelength range of FDML lasers to the 1060nm spectral region[29, 30] enabled high quality ultra-fast retinal swept source OCT imaging with good penetration into the choroid[7, 29, 31]. However, up to now, FDML showed inferior coherence compared to the most modern VCSEL based OCT lasers. So we demonstrate a new, very low noise FDML laser and demonstrate its meter class instantaneous coherence properties.

## 2. METHOD

We integrated the entire FDML fiber cavity into a 3U tall, temperature controlled 19 inch rack enclosure. All main components (SOA, FFP-TF, cFBG, spool) have their own active temperature regulation and are partly (SOA, spool) thermally isolated from each other. The whole cavity is an 8:1 mix of Corning HI1060 and SMF28e fiber. The use of HI1060 fiber in the delay spool makes the cavity dispersion monotonic to enable simple dispersion compensation with a single custom made chirped fiber Bragg grating (cFBG from Teraxion). To reduce the overall cavity losses, the cFBG is also used as an output coupler. The dispersion of this cFBG can be tuned by applying a temperature gradient along the fiber, its holder is in contact with one Peltier element at each end. The home built fiber Fabry Perot tunable filter (FFP) runs resonantly at 411 kHz. The SOA is modulated to operate at 1/8 duty cycle to allow 8 times optical buffering for a resulting wavelength sweep repetition rate of 3.28MHz. For OCT imaging a buffer stage, booster, interferometer and detection unit described in an earlier publication[32] were used. The output power at the sample arm was 40 mW.

## 3. RESULTS

The remaining group delay differences in the cavity throughout the sweep range were difficult to measure but we estimate them to be below 200 fs throughout the full sweep range. We acquired intensity traces with a 70 GHz real time oscilloscope and a 50 GHz photodiode to reveal the effect of intra cavity dispersion. This was simulated by detuning the laser from its optimum frequency by 1 Hz. As can be seen in Fig 1. the trace then is full of "cracks" which had been identified as Nozaki-Bekki holes[33], where the intensity drops close to zero, followed by spikes which reach up to double the mean intensity level.

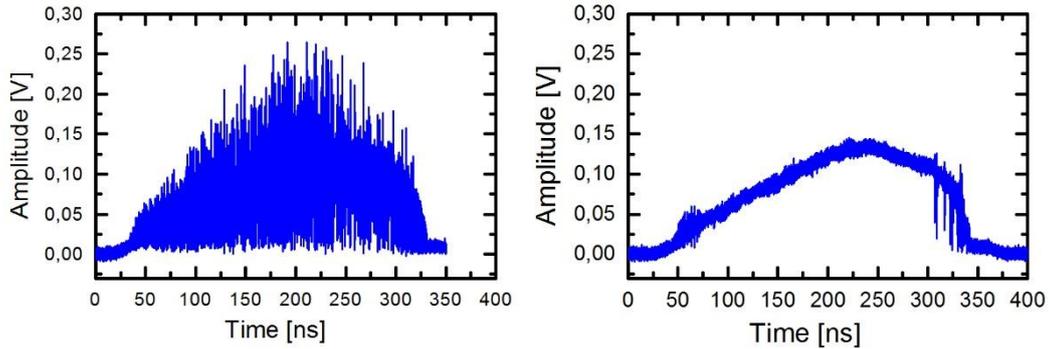

Fig. 1 Backward (blue to red) 130nm sweep intensity trace of the highly dispersion compensated but detuned (1Hz) laser (left) and Intensity trace of ultra low noise operation (right) acquired with 50 GHz analog bandwidth.

When the FFP frequency is adjusted down to a few mHz, the cracks almost disappear. Except for a small region on the red end the noise is completely dominated by the photodetector noise. This very low noise operation mode has been observed before[34, 35] but only over a very small spectral range of 3.5 nm. To exclude the noisy region at the red end of the sweeps, the sweep range was limited to 120 nm in the following.

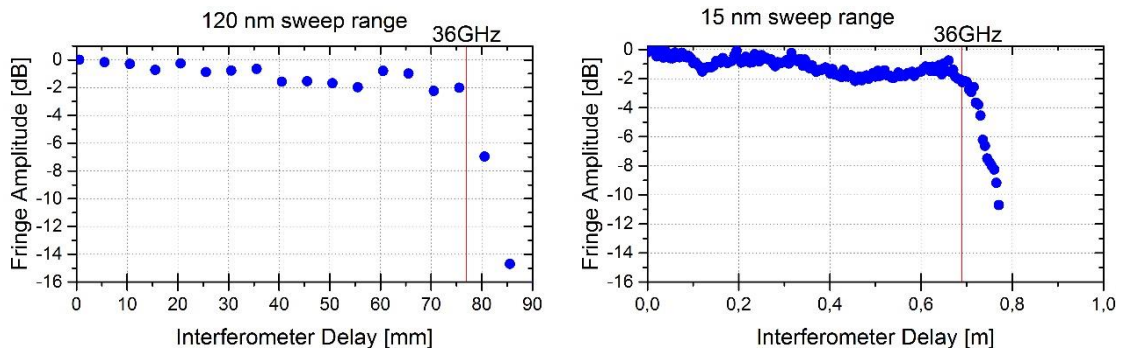

Fig. 2 Fringe amplitude decay for very long imaging ranges of a 120 nm (left) and 15 nm sweep (right). The fringe amplitude goes down by less than 3 dB before the fringe frequencies exceed the 36 GHz detection bandwidth (red line) of the real time oscilloscope.

To evaluate the lasers suitability for long range optical coherence tomography we measured the decrease of the interference fringe amplitude for long optical path length differences in a Mach-Zehnder interferometer. We measured this amplitude roll-off instead of a real OCT signal roll-off, because the measurement was performed on a single oscilloscope channel which could not be k-clocked. Here, we used a 36 GHz bandwidth oscilloscope. The fringe amplitude was measured in the center of the sweep. Because the fringe frequencies exceeded the analog detection bandwidth of our measurement system for high interferometer delays, we also performed a measurement with a narrower laser sweep range (15 nm) resulting in reduced fringe frequencies. With this setting we saw nice interference-fringes up to delays of 70 cm where the fringe frequencies again exceeded the detection bandwidth of the oscilloscope.

As can be seen in Fig. 2 where the fringe amplitude is plotted against the interferometer delay, the fringe amplitude decreases by less than 3 dB until the fringe frequency reaches the limit of our detection bandwidth. We emphasize that this amplitude roll-off does not directly translate to the signal roll-off commonly referred to in OCT as it is challenging to find a suitable method for the k-linearization of such high frequency sweeps.

**Long-range imaging with the new laser**

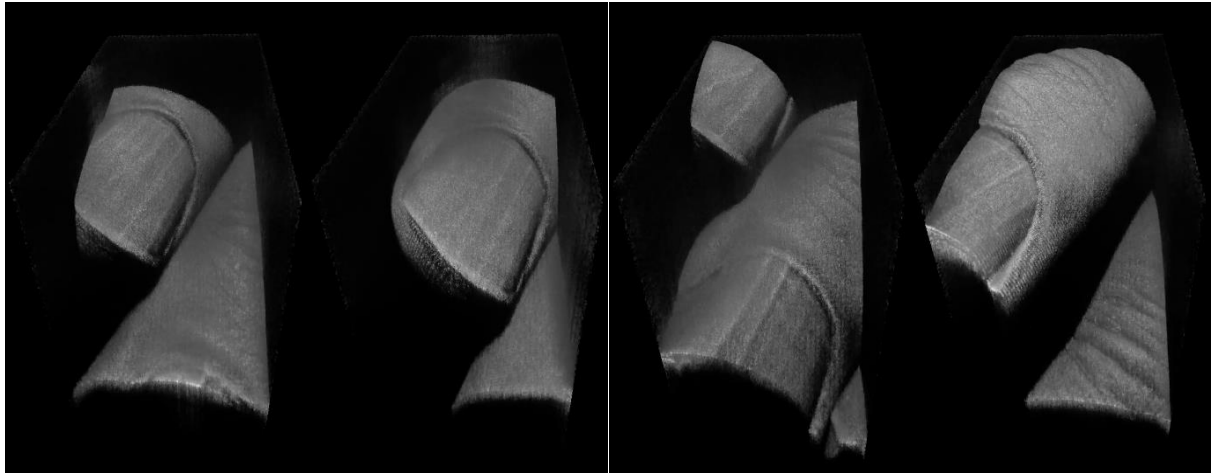

Fig. 3 Images taken from a 4D OCT video that was acquired and processed and displayed in real time using the Optores GPU OCT processing library. The imaging range, limited by the 1.9 GHz detection bandwidth of the digitizer card, was set to 25 mm by narrowing the laser sweep range to 22 nm.

To demonstrate the immediate benefit of its improved coherence, we decided to increase the axial imaging range of our OCT system not by increasing the acquisition bandwidth but by narrowing down the sweep range of the laser as it has been demonstrated before[18]. While the increased imaging range then comes at the cost of axial resolution, making the visualization of subsurface structures difficult, this technique has the advantage that it can be directly employed in a system for live 4D-OCT imaging which has been demonstrated before[6]. Detection bandwidth, ADC streaming capability and processing power requirements are unchanged while axial resolution is traded in for imaging range, potentially enabling seamless zooming in future live 4D OCT implementations.

We used a 4 GS AlazarTech card and reduced the sweep range of the laser to 22 nm to enhance the resulting imaging range to 2.5 cm, which was the maximum 1.6 GHz fringe frequency that our system could handle. In combination with a 2.7 kHz resonant scanning galvanometer, we acquired, processed and rendered 22 Volumes per second, 240 frames with 300 A-scans each in real time live with less than 1 volume lag in display. Fig. 3 shows images of finger tips, taken from a recorded video. Comparing this video to previously published real time 4D OCT videos[6], the image quality is clearly improved even though the imaging range as well as the A-scan rate have been increased substantially.

To further extend the imaging range, we also ran the laser at ~4 nm sweep range resulting in approximately 10 cm of axial imaging range and implemented post-objective scanning with a 1000 mm lens. Here, four seconds of raw data were acquired and stored in the onboard memory of the 4 GS data acquisition card and post-processed later on. Within these four seconds, 44 volumes were acquired, resulting in an effective volume rate of 11 volumes per second. Each volume was made up of 480 x 480 A-scans. In this configuration, probably due to the very low NA, the signal levels were quite low. However, after averaging the a-scans in groups of four by four the image quality is good enough to produce high quality renderings of a human face, as can be seen in **Fehler! Verweisquelle konnte nicht gefunden werden.**.

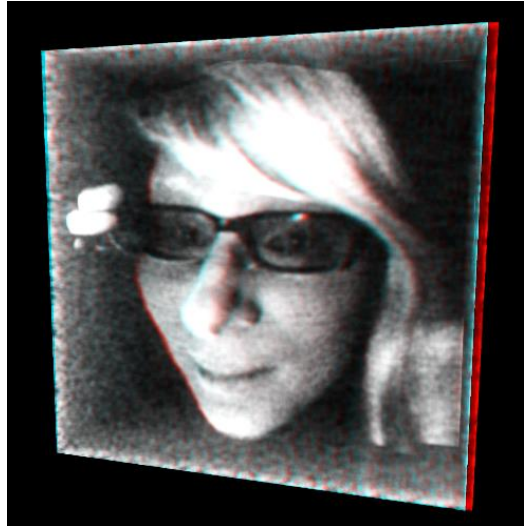

Fig. 4 Anaglyph rendered 3D image of a human face acquired in < 100 ms

## 4. CONCLUSION

Here we demonstrated our implementation of an FDML laser operating in an ultra-low noise sweet spot mode throughout a 120 nm sweep range at a center wavelength of 1292 nm. We observe meter class instantaneous coherence properties considering fringe visibility. We show the improvement in 3D OCT image quality, rendering 22 volumes per second live in real time with 25 mm of imaging range. We also demonstrate volumetric OCT of a moving human face with 11 volumes per second and 10 cm of imaging range.